\documentclass{article}

\usepackage{booktabs}
\usepackage{array}
\usepackage{lscape}
\usepackage{threeparttable}
\usepackage{booktabs}
\usepackage{array}
\usepackage{multirow}
\usepackage{arxiv}
\usepackage[utf8]{inputenc}
\usepackage[T1]{fontenc}
\usepackage{hyperref}
\usepackage{url}
\usepackage{booktabs}
\usepackage{amsfonts}
\usepackage{nicefrac}
\usepackage{microtype}
\usepackage{graphicx}
\graphicspath{{./images/}}

\title{Impact of Clinical Decision Support Systems (CDSS) on Clinical Outcomes and Healthcare Delivery in Low- and Middle-Income Countries: Protocol for a Systematic Review and Meta-analysis}

\author{
Garima Jain$^{1}$ \\
\texttt{garima.jain@icmr.gov.in}
\And
Anand Bodade$^{2}$ \\
\texttt{anand.bodade@icmr.gov.in}
\And
Sanghamitra Pati$^{1,2,3}$ \\
\texttt{sanghamitra.pati@icmr.gov.in}
\\[1em]
$^{1}$ICMR National Institute for Research in Digital Health \& Data Science, New Delhi, India \\
$^{2}$ICMR Headquarters, New Delhi, India \\
$^{3}$ICMR-RMRC Bhubaneshwar, Odisha, India
}

\begin{document}
\maketitle

\begin{abstract}
\subsection*{Introduction}
Clinical decision support systems (CDSS) are used to improve clinical and service outcomes, yet evidence from low- and middle-income countries (LMICs) is dispersed. This protocol outlines methods to quantify the impact of CDSS on patient and healthcare delivery outcomes in LMICs.
\subsection*{Methods and analysis}
We will include comparative quantitative designs (randomised trials, controlled before–after, interrupted time series, comparative cohorts) evaluating CDSS in World Bank–defined LMICs. Standalone qualitative studies are excluded; mixed-methods studies are eligible only if they report comparative quantitative outcomes, for which we will extract the quantitative component. Searches (from inception to 30 September 2024) will cover MEDLINE, Embase, CINAHL, CENTRAL, Web of Science, Global Health, Scopus, IEEE Xplore, LILACS, African Index Medicus, and IndMED, plus grey sources. Screening and extraction will be performed in duplicate. Risk of bias will be assessed with RoB 2 (randomised trials) and ROBINS-I (non-randomised). Random-effects meta-analysis will be performed where outcomes are conceptually/statistically comparable; otherwise, a structured narrative synthesis will be presented. Heterogeneity will be explored using $I^{2}/\tau^{2}$ and a-priori subgroups/meta-regression (condition area, care level, CDSS type,
readiness proxies, study design).

\end{abstract}


\section*{Ethics and dissemination}
This review uses published data and does not require ethics approval. Results will be disseminated via peer-reviewed publication, conference presentations, and LMIC-oriented policy briefs; extraction templates and analysis code will be shared openly on publication.

\section*{Strengths and Limitations of the Study}
\begin{itemize}
    \item The protocol is registered (PROSPERO CRD42024599329) and follows PRISMA/PRISMA-P guidance.
    \item We use a multi-database search (biomedical + technical sources) with reference-list screening to maximise retrieval.
    \item Dual independent screening and data extraction are planned, with design-appropriate risk-of-bias appraisal (RoB 2 for RCTs; JBI checklists for observational designs).
    \item A pre-specified quantitative synthesis plan uses random-effects meta-analysis where outcomes are comparable, with subgroup and sensitivity analyses and narrative synthesis otherwise.
    \item Limitation: restricting to English full texts and the last 10 years may introduce language/time-window bias and under-represent LMIC evidence.
\end{itemize}

\section{Background}
Clinical Decision Support Systems (CDSS) are advanced technological tools designed to enhance healthcare decision-making by integrating clinical knowledge with patient data to provide personalized recommendations. CDSS have been widely implemented to aid healthcare providers in clinical decision-making, leveraging digital technologies to improve diagnostic accuracy, guide treatment decisions, and optimize healthcare outcomes.\cite{cite1_benimetskaya2023retrospective, cite2_agarwal2021decision, cite3_arseniev2020cerebral} These are an integral part of health information technologies and are often part of electronic health record (EHR) systems to provide real-time, evidence-based recommendations to clinicians during patient care.\cite{cite4_lewkowicz2020economic}

The origin of CDSS dates back to the late 1960s when the first rule-based systems, such as MYCIN, were developed to assist in diagnosing bacterial infections and suggesting appropriate treatments.\cite{cite5_chen2023harnessing} Over time, these systems have evolved considerably, incorporating more sophisticated technologies, including artificial intelligence (AI), machine learning (ML), and natural language processing (NLP). Modern CDSS can analyse vast and complex datasets, enabling healthcare providers to make more informed decisions, whether it be diagnosing a disease, selecting the most appropriate treatment plan, or predicting patient outcomes based on clinical history and available medical data.

CDSS encompasses a variety of tools, including computerized alerts and reminders, clinical guidelines, condition-specific order sets, focused patient data reports, diagnostic support, and knowledge-based systems that guide treatment decisions. These systems help address common issues in healthcare delivery, such as the need for timely decision-making, the overload of clinical data, and variability in clinical practice. CDSS can positively influence patient-level outcomes such as reducing diagnostic errors, improving medication safety, and enhancing treatment effectiveness, while also improving healthcare delivery outcomes like patient throughput, resource allocation, and overall care quality.\cite{cite1_benimetskaya2023retrospective, cite5_chen2023harnessing, cite6_alcorn2022evaluation, cite7_akay2023artificial, cite8_albin2022development}

CDSS applications are diverse e.g. in the management of chronic diseases like diabetes, CDSS can suggest appropriate interventions based on blood glucose levels and other clinical parameters. In oncology, CDSS assists in treatment planning by integrating genomic data to recommend targeted therapies. In emergency care, CDSS aids in triage systems, helping clinicians quickly assess patients based on their symptoms and history, ensuring rapid response and intervention. In mental health, CDSS can suggest probable diagnoses and treatment options based on patient-reported outcomes and clinical data, allowing for personalized care strategies.

CDSS has a potential to transform healthcare. By integrating knowledge from clinical guidelines, research, and patient data, these systems provide decision support that is not only evidence-base  but tailored to individual patient needs also. e.g. in acute stroke management, it can predict patient outcomes and guide interventions, which can lead to improved survival rates and reduced healthcare costs.\cite{cite7_akay2023artificial} Moreover, the dynamic decision-making capabilities of CDSS where the system can update itself with new data and guidelines, allow for continuous improvements in diagnostic accuracy and treatment efficacy, making it an invaluable tool in evolving clinical environments.

There are few challenges in application of CDSS. E.g. with frequent system alerts clinician can become desensitized and this can reduce effectiveness of system. CDSS must be carefully integrated into clinical workflows to avoid disruptions in medical care. Sometimes, complex interfaces can lead to user frustration and hinder clinical adoption. Interoperability with existing EHRs and healthcare systems is essential to ensure seamless data exchange and system functionality. For successful implementation of CDSS alignment with local clinical guidelines and healthcare infrastructure is important.

In high-income countries, CDSS are popular owing to their impact on improving healthcare delivery. AI-driven CDSS systems have shown improved diagnostic accuracy and decision-making in fields such as cardiology, oncology, and emergency care. These systems also reduce the cognitive load on clinicians by providing automated, evidence-based recommendations, allowing for more consistent and timely care, reducing medication errors, decreasing unnecessary diagnostic testing, and enhancing overall patient safety and care. CDSS in low- and middle-income countries (LMICs) have potential to improve clinical outcomes. electronic decision algorithms, such as e-ALMANACH, have been used to manage febrile illnesses in children, resulting in improved clinical outcomes and more rational use of antibiotics. These systems can guide healthcare workers through decision-making processes, reducing clinical failure rates and unnecessary treatments. However, despite these benefits, the use of CDSS in LMICs remains underutilized, largely due to a lack of awareness, training, and infrastructure.

Implementation of CDSS in low- and middle-income countries (LMICs) is challenging due to poor infrastructure, scarce healthcare personnel, resource constraints.\cite{cite9_witter2022learning} There are barriers in adopting CDSS like . limited access to digital health infrastructure, lack of technical expertise, and financial constraints.\cite{cite10_tcheng2017optimizing} It should be noted that most of CDSS are developed considering their use in high-resource settings, making them less applicable to the healthcare realities in LMICs, where diagnostic tools and clinical guidelines may differ significantly.

Most of the existing research focuses on high-income countries, leaving a critical need for evaluating the effectiveness of CDSS in resource-poor nations. Understanding the barriers in their implementation can help to strengthen their applications in these countries. . Also, evidence on the scalability and sustainability of CDSS in these regions is limited, highlighting the need for more research to explore how these systems can be adapted to fit the needs of LMICs. Current Systematic Review is planned to understand the effect of using CDSS in LMICs and to bridge the gap in the evidence based clinical management especially in LMICs and possible measures that can be considered for their use at best possible ways.

\section{Methods and Design}
We will follow the PRISMA guidelines. The protocol has been registered with PROSPERO (CRD42024599329). Studies will be identified through searches of the following databases: PubMed, Embase, Cochrane Library, Medline, IEEE Explore, and Google Scholar. Study dates. Searches will cover database inception to 30 June 2025; screening/extraction will occur July–September 2025 with analysis and submission of results targeted for Q1 2026.

\subsection{Eligibility Criteria and study selection (Table \ref{table1})}
\begin{itemize}
    \item Inclusion Criteria
    \begin{itemize}
            \item Studies involving digital technology (including, but not limited to AI) as the intervention.
            \item Knowledge and non-knowledge-based CDSS.
            \item Cohort, cross-sectional, prospective observational, interventional studies, and randomized control trials.
            \item Language: Studies must be published in English. Records in any language will be searched and screened at title/abstract level; however, only English full texts will be assessed for inclusion. We will record the number of non-English full texts excluded at eligibility and summarise their key characteristics (e.g., country, condition area) when available from the abstract. Searches will be executed without language limits to map potentially relevant non-English literature; eligibility will apply the English-only full-text criterion.
            \item Date of publication: Search will cover articles published in the last ten years in the databases up to 30th September 2024. 
    \end{itemize}
    \item Exclusion criteria
    \begin{itemize}
        \item Studies limited to developing/proposing models or frameworks.
        \item Studies not reporting patient-level outcomes or healthcare workflow outcomes.
        \item Non-peer-reviewed or qualitative reviews, viewpoints, editorials, abstracts, conference papers, treatment guidelines, and prevalence studies. Standalone qualitative studies are excluded; mixed-methods studies are eligible only if they report comparative quantitative outcomes, in which case we will use the quantitative component.
    \end{itemize}
    \item \textbf{Population (P):} Patients, healthcare personnel, or healthcare systems in low and middle income countries.\cite{cite11_worldbank_country_lending_2024}
    \item \textbf{Intervention (I):} Implementation or use of CDSS that consist of at least one digital technology (including, but not limited to artificial intelligence).
    \item \textbf{Comparison (C):} Healthcare systems with CDSS versus those without CDSS or traditional decision-making processes.
    \item \textbf{Outcomes (O): Primary Outcome}
    \begin{itemize}
        \item Patient outcomes (e.g., diagnostic accuracy, time to diagnosis, treatment decisions, adherence to treatment guidelines, self-management, mortality reduction etc).
        \item Healthcare workflow outcomes (e.g., efficiency, quality, burden, delivery outcomes).
    \end{itemize}
    \item \textbf{Search Strategy:} The search will be performed using a combination of index terms, including 'clinical decision support system', 'CDSS', and 'healthcare delivery' across the databases mentioned above. (Table \ref{table2}) Additional studies will be identified by examining reference lists of reviews and eligible publications.
\end{itemize}

\subsection{Risk of Bias assessment}
To evaluate the risk of bias in included studies, pairs of reviewers will independently assess each study using design-appropriate, current tools. For randomised controlled trials, we will use the Cochrane Risk of Bias 2 (RoB 2) tool, which assesses bias across five domains: randomisation process, deviations from intended interventions, missing outcome data, outcome measurement, and selection of the reported result. Both reviewers will apply RoB 2 at the outcome level following the official guidance; disagreements will be resolved by discussion or a third reviewer \cite{cite12_higgins2011cochrane}. For non-randomised comparative studies (e.g. cohort, case-control, controlled before–after, interrupted time series), we will use the ROBINS-I tool, which evaluates bias due to confounding, participant selection, intervention classification, deviations from intended interventions, missing data, outcome measurement, and selective reporting.

Where available, we will provide visual summaries (traffic-light plots) of risk-of-bias judgments. Sensitivity analyses will exclude studies judged to be at high or critical risk of bias to assess the robustness of pooled estimates \cite{cite13_munn2020methodological}. In cases of disagreement between the reviewers, discussions will be held to reach a consensus, with input from a third reviewer if required. The results of the risk-of-bias assessments, as well as the extracted data, will be entered into a standardized sheet for further synthesis and statistical analysis

\subsection{Data Extraction}
Data extraction will be independently conducted by two reviewers. Discrepancies will be resolved through discussion. For our study, we will utilize a standardized data abstraction form accompanied by a comprehensive instruction manual. To ensure consistency and accuracy in data extraction, calibration exercises will be conducted prior to the extraction process. Two reviewers, working in pairs, will independently extract the following information from the eligible studies: (i) study characteristics, including the names of authors, publication year, country of origin, and funding sources; (ii) population-related details such as the number of participants, average age, gender distribution, the type of clinical condition or disease studied, and duration of follow-up; (iii) intervention and comparator details, such as the type of intervention, mode of delivery, duration of treatment, and timing of interventions; and (iv) the outcomes of interest related to clinical decision support systems, including clinical outcomes, healthcare delivery outcomes, and any adverse events reported. A meta-analysis will be conducted if sufficient homogeneous data are available; otherwise, a narrative synthesis will be performed.

\subsection{Data synthesis}
Data synthesis will follow a structured approach to combine and analyse findings from the included studies. We will first perform a narrative synthesis for all studies, summarizing the characteristics, interventions, outcomes, and key findings of each study. This will provide a broad overview of the evidence.

For quantitative data, we will perform a meta-analysis if the studies are sufficiently homogeneous in terms of population, interventions, and outcomes. This will involve pooling effect sizes using a random-effects model, which accounts for variations between studies. For dichotomous outcomes, such as clinical outcomes or adverse events, we will calculate relative risks (RR) with 95\% confidence intervals (CI). For continuous outcomes, such as treatment effects or health improvements, we will use mean differences (MD) or standardized mean differences (SMD) with 95\% CI depending on the nature of the data.

Heterogeneity among studies will be assessed using the Cochran’s Q test and $I^{2}$ statistics, with significant heterogeneity explored further through subgroup analyses and sensitivity analyses. Subgroup analyses may include factors such as population characteristics, setting (e.g., rural vs. urban), and type of intervention (e.g., provider-led vs. self-managed interventions). Sensitivity analyses will be conducted to test the robustness of the findings, particularly by excluding studies with a high risk of bias or outlier results.

Where meta-analysis is not possible due to significant heterogeneity or lack of sufficient comparable data, we will present a descriptive synthesis. In such cases, we will synthesize results narratively, emphasizing trends and patterns in the data, and using tables and figures for visual representation. Additionally, we will use tools such as forest plots to illustrate effect sizes and funnel plots to assess potential publication bias. By using these techniques, our data synthesis will provide a comprehensive evaluation of the impact of CDSS in the specific settings of LMICs.

\subsection{Assessing the Quality of Evidence \cite{cite14_bezerra2022assessment}}
We will use the Grading of Recommendations Assessment, Development, and Evaluation (GRADE) (Table \ref{table3}) framework to systematically assess the quality of evidence in our study. GRADE will help evaluate the certainty of the evidence across different domains including study design, risk of bias, consistency, directness, and precision. Evidence from randomized controlled trials (RCTs) will initially be considered as high quality, but may be downgraded based on identified limitations such as high risk of bias, inconsistency in results, or imprecision. For non-randomized studies, the starting certainty will be lower, and we will assess the risk of bias using tools like the ROBINS-I. The GRADE framework will also enable us to consider the overall strength of recommendations, particularly in resource-limited settings where evidence might be sparse. By using GRADE, we ensure that the conclusions drawn from our review are transparent and based on the best available evidence, providing clear guidance for both clinical practice and policy-making

\textbf{Patient and public involvement.} Patients or the public were not involved in the design, conduct, reporting, or dissemination plans of this research.

\textbf{Ethics and dissemination}. This study involves secondary analysis of published data and does not require ethics approval. Findings will be disseminated via journal publication, conference presentations, and LMIC policy briefs; upon publication, de-identified extraction sheets and analytic code will be made available in an open repository.

\section{Discussion}
Clinical Decision Support Systems (CDSS) have become an integral part of modern healthcare, particularly in high-income countries where digital infrastructure is more robust.\cite{cite15_kumar2022extra, cite16_jain2020peripheral, cite17_jain2019esophageal} The potential of CDSS to improve clinical outcomes, enhance decision-making, and streamline healthcare workflows is well-documented. However, the evidence supporting its implementation in low- and middle-income countries (LMICs) is limited. This systematic review aims to address this gap by assessing the impact of CDSS in resource-poor settings and providing insights into its effectiveness, scalability, and adaptability.

The review aims to evaluate the impact of CDSS on healthcare systems in resource-poor settings, potentially guiding future implementations and helping healthcare providers and policymakers develop strategies tailored to the unique challenges of LMICs. Our study offers several strengths compared to existing reviews. First, we will comprehensively evaluate the comparative effectiveness of CDSS in a variety of healthcare settings, particularly those in resource-constrained environments. This focus on LMICs will provide valuable insights into how these systems can be adapted to fit local healthcare needs. Second, by utilizing the GRADE approach, we will systematically assess the quality of evidence across all included studies, ensuring that the conclusions drawn are based on the highest possible quality of evidence. This will enhance the reliability of our findings and their applicability in guiding healthcare policies and decision-making processes in LMICs.

The potential impact and strengths of this review extends beyond providing evidence on CDSS implementation. By synthesizing the existing data, we aim to highlight the areas where CDSS can significantly contribute to improving healthcare outcomes, particularly in managing chronic diseases, enhancing diagnostic accuracy, and supporting clinical workflows. Anticipated challenges and limitations of this study include the availability and diversity of the studies needed to perform robust data synthesis. The limited number of high-quality studies focusing on CDSS in LMICs might pose a constraint. Additionally, the heterogeneity of healthcare settings, infrastructure, and population characteristics across LMICs may limit the generalizability of the findings. These factors will need to be carefully addressed through subgroup analyses and sensitivity analyses during data synthesis.

Despite these challenges, this review has the potential to significantly contribute to the understanding of CDSS in resource-limited settings. By highlighting both the benefits and the barriers to implementation, this study will provide crucial insights for future research and policy development. Moreover, the findings will be disseminated widely through peer-reviewed publications and presentations at national and international conferences to ensure broad accessibility and impact, thereby helping bridge the knowledge gap on CDSS in LMICs and guiding future efforts to optimize healthcare delivery in these regions. We used the PRISMA-P checklist when writing our report.\cite{cite18_moher2015preferred}

\begin{table}[htbp]
\centering
\caption{Inclusion and Exclusion Criteria for Study Selection}
\label{table1}
\begin{tabular}{p{2.5cm} p{6cm} p{6cm}}
\toprule
\textbf{Category} & \textbf{Inclusion Criteria} & \textbf{Exclusion Criteria} \\
\midrule

\textbf{Population} &
Patients, healthcare personnel, or healthcare systems in low and middle income countries &
Studies focusing only on high-income countries \\

\midrule
\textbf{Intervention} &
Implementation or use of clinical decision support systems (CDSS) utilizing digital technology, including AI-based tools for healthcare delivery and decision support &
Studies focused on the development of CDSS models without real-world implementation, or those limited to framework proposals \\

\midrule
\textbf{Comparator} &
Healthcare systems using CDSS compared to systems without CDSS, or traditional decision-making processes without digital support &
Absence of a control group or inappropriate comparator \\

\midrule
\textbf{Outcome} &
Primary: Patient outcomes (diagnostic accuracy, treatment adherence, improved clinical outcomes) \newline
Secondary: Healthcare delivery outcomes (efficiency, workflow optimization, quality of care) &
Studies using non-validated methods to measure the outcomes, or lacking patient-level or healthcare workflow outcomes \\

\midrule
\textbf{Study Design} &
Randomized controlled trials (RCTs), cohort studies, prospective interventional studies &
Non-randomized controlled trials, qualitative studies, reviews, commentaries, and non-peer-reviewed articles \\

\bottomrule
\end{tabular}
\end{table}

\begin{table}[htbp]
\centering
\caption{Search strategy for PubMed}
\label{table2}
\begin{tabular}{p{16cm}}
\toprule
\textbf{Category} : \textbf{Search Terms} \\
\midrule

\textbf{Population :} 
('Afghanistan' [Title/Abstract] OR 'Albania' [Title/Abstract] OR 'Algeria' [Title/Abstract] OR 'Angola' [Title/Abstract] OR 'Argentina' [Title/Abstract] OR 'Armenia' [Title/Abstract] OR 'Azerbaijan' [Title/Abstract] OR 'Bangladesh' [Title/Abstract] OR 'Belarus' [Title/Abstract] OR 'Belize' [Title/Abstract] OR 'Benin' [Title/Abstract] OR 'Bhutan' [Title/Abstract] OR 'Bolivia' [Title/Abstract] OR 'Bosnia and Herzegovina' [Title/Abstract] OR 'Botswana' [Title/Abstract] OR 'Brazil' [Title/Abstract] OR 'Burkina Faso' [Title/Abstract] OR 'Burundi' [Title/Abstract] OR 'Cabo Verde' [Title/Abstract] OR 'Cambodia' [Title/Abstract] OR 'Cameroon' [Title/Abstract] OR 'Central African Republic' [Title/Abstract] OR 'Chad' [Title/Abstract] OR 'China*' [Title/Abstract] OR 'Colombia' [Title/Abstract] OR 'Comoros' [Title/Abstract] OR 'Democratic Republic of Congo' [Title/Abstract] OR 'Congo' [Title/Abstract] OR 'Costa Rica' [Title/Abstract] OR 'Cote d'Ivoire' [Title/Abstract] OR 'Cuba' [Title/Abstract] OR 'Djibouti' [Title/Abstract] OR 'Dominica' [Title/Abstract] OR 'Dominican Republic' [Title/Abstract] OR 'Ecuador' [Title/Abstract] OR 'Egypt' [Title/Abstract] OR 'El Salvador' [Title/Abstract] OR 'Equatorial Guinea' [Title/Abstract] OR 'Eritrea' [Title/Abstract] OR 'Eswatini' [Title/Abstract] OR 'Ethiopia' [Title/Abstract] OR 'Fiji' [Title/Abstract] OR 'Gabon' [Title/Abstract] OR 'Gambia' [Title/Abstract] OR 'Georgia' [Title/Abstract] OR 'Ghana' [Title/Abstract] OR 'Grenada' [Title/Abstract] OR 'Guatemala' [Title/Abstract] OR 'Guinea' [Title/Abstract] OR 'Guinea-Bissau' [Title/Abstract] OR 'Guyana' [Title/Abstract] OR 'Haiti' [Title/Abstract] OR 'Honduras' [Title/Abstract] OR 'India' [Title/Abstract] OR 'Indonesia' [Title/Abstract] OR 'Iran' [Title/Abstract] OR 'Iraq' [Title/Abstract] OR 'Jamaica' [Title/Abstract] OR 'Jordan' [Title/Abstract] OR 'Kazakhstan' [Title/Abstract] OR 'Kenya' [Title/Abstract] OR 'Kiribati' [Title/Abstract] OR 'Democratic People's Republic of Korea' [Title/Abstract] OR 'Kosovo' [Title/Abstract] OR 'Kyrgyzstan' [Title/Abstract] OR 'Lao People's Democratic Republic' [Title/Abstract] OR 'Lebanon' [Title/Abstract] OR 'Lesotho' [Title/Abstract] OR 'Liberia' [Title/Abstract] OR 'Libya' [Title/Abstract] OR 'North Macedonia' [Title/Abstract] OR 'Madagascar' [Title/Abstract] OR 'Malawi' [Title/Abstract] OR 'Malaysia' [Title/Abstract] OR 'Maldives' [Title/Abstract] OR 'Mali' [Title/Abstract] OR 'Marshall Islands' [Title/Abstract] OR 'Mauritania' [Title/Abstract] OR 'Mauritius' [Title/Abstract] OR 'Mexico' [Title/Abstract] OR 'Micronesia' [Title/Abstract] OR 'Moldova' [Title/Abstract] OR 'Mongolia' [Title/Abstract] OR 'Montenegro' [Title/Abstract] OR 'Montserrat' [Title/Abstract] OR 'Morocco' [Title/Abstract] OR 'Mozambique' [Title/Abstract] OR 'Myanmar' [Title/Abstract] OR 'Namibia' [Title/Abstract] OR 'Nauru' [Title/Abstract] OR 'Nepal' [Title/Abstract] OR 'Nicaragua' [Title/Abstract] OR 'Niger' [Title/Abstract] OR 'Nigeria' [Title/Abstract] OR 'Niue' [Title/Abstract] OR 'Pakistan' [Title/Abstract] OR 'Panama' [Title/Abstract] OR 'Papua New Guinea' [Title/Abstract] OR 'Paraguay' [Title/Abstract] OR 'Peru' [Title/Abstract] OR 'Philippines' [Title/Abstract] OR 'Rwanda' [Title/Abstract] OR 'Saint Helena' [Title/Abstract] OR 'Samoa' [Title/Abstract] OR 'Sao Tome and Principe' [Title/Abstract] OR 'Senegal' [Title/Abstract] OR 'Serbia' [Title/Abstract] OR 'Sierra Leone' [Title/Abstract] OR 'Solomon Islands' [Title/Abstract] OR 'Somalia' [Title/Abstract] OR 'South Africa' [Title/Abstract] OR 'South Sudan' [Title/Abstract] OR 'Sri Lanka' [Title/Abstract] OR 'Saint Lucia' [Title/Abstract] OR 'Saint Vincent and the Grenadines' [Title/Abstract] OR 'Sudan' [Title/Abstract] OR 'Suriname' [Title/Abstract] OR 'Syrian Arab Republic' [Title/Abstract] OR 'Tajikistan' [Title/Abstract] OR 'Tanzania' [Title/Abstract] OR 'Thailand' [Title/Abstract] OR 'Timor-Leste' [Title/Abstract] OR 'Togo' [Title/Abstract] OR 'Tokelau' [Title/Abstract] OR 'Tonga' [Title/Abstract] OR 'Tunisia' [Title/Abstract] OR 'Turkey' [Title/Abstract] OR 'Turkmenistan' [Title/Abstract] OR 'Tuvalu' [Title/Abstract] OR 'Uganda' [Title/Abstract] OR 'Ukraine' [Title/Abstract] OR 'Uzbekistan' [Title/Abstract] OR 'Vanuatu' [Title/Abstract] OR 'Venezuela' [Title/Abstract] OR 'Vietnam' [Title/Abstract] OR 'Wallis and Futuna' [Title/Abstract] OR 'West Bank and Gaza Strip' [Title/Abstract] OR 'Yemen' [Title/Abstract] OR 'Zambia' [Title/Abstract] OR 'Zimbabwe' [Title/Abstract] OR 'LMIC'[Title/Abstract] OR 'low resource'[Title/Abstract] OR 'resource-poor'[Title/Abstract] OR 'low income countr*'[Title/Abstract] OR 'low-middle income countr*'[Title/Abstract] OR 'low-income countr'[Title/Abstract]\\

\midrule
\textbf{Intervention:}  
('clinical decision support system’[Title/Abstract] OR 'clinical decision support'[Title/Abstract] OR ‘Electronic decision support system’[Title/Abstract] OR ‘computeri* clinical decision support system’[Title/Abstract] OR 'decision support system'[Title/Abstract]) OR ('CDSS'[Title/Abstract]) OR ('decision support tool'[Title/Abstract]) \\

\midrule
\textbf{Outcome:}
('outcome*'[Title/Abstract] OR ‘management*'[Title/Abstract] OR ‘intervention*’[Title/Abstract] OR ‘diagnos*’[Title/Abstract] OR ‘treatment*’[Title/Abstract] OR ‘guideline* adherence’[Title/Abstract] OR 'alert*' [Title/Abstract] OR ‘healthcare delivery'[Title/Abstract] OR 'quality of care'[Title/Abstract] OR 'quality of healthcare'[Title/Abstract] OR ‘universal health coverage’[Title/Abstract] OR ‘healthcare infrastructure’[Title/Abstract]' OR cost*' OR 'cost-effectiveness'[Title/Abstract] OR ‘mortality reduc*’ [Title/Abstract] OR ‘healthcare efficiency’[Title/Abstract] OR ‘healthcare quality’[Title/Abstract] OR ‘burden’[Title/Abstract] OR ‘information management’[Title/Abstract] OR ‘ insurance*’[Title/Abstract] OR ‘health record*’ [Title/Abstract] OR ‘discharg*’ [Title/Abstract] OR ‘health*’ [Title/Abstract] OR ‘follow-up’ [Title/Abstract] OR ‘administrat*’ [Title/Abstract] OR ‘satisf*’ [Title/Abstract] OR ‘healthcare workflow’[Title/Abstract] OR ‘patient workflow’[Title/Abstract] OR ‘primary care’[Title/Abstract] OR ‘integrated care’[Title/Abstract]) \\

\midrule
\textbf{Study Design:}
("randomized controlled trial"[Title/Abstract] OR "RCT"[Title/Abstract] OR "cohort study"[Title/Abstract] OR "interventional study"[Title/Abstract]) \\

\bottomrule
\end{tabular}
\end{table}

\begin{landscape}
\begin{table}[htbp]
\centering
\caption{GRADE Evidence Profile of the Impact of CDSS on Healthcare Outcomes in Low and Middle Income Countries}
\label{table3}
\begin{threeparttable}
\begin{tabular}{p{4cm} p{3cm} p{1.8cm} p{2cm} p{1.8cm} p{1.8cm} p{2cm} p{2.5cm} p{2.8cm}}
\toprule
\textbf{Outcome} & 
\textbf{No. of studies (design)} & 
\textbf{Risk of bias} & 
\textbf{Inconsistency} & 
\textbf{Indirectness} & 
\textbf{Imprecision} & 
\textbf{Publication bias} & 
\textbf{Effect size (95\% CI)} & 
\textbf{Overall certainty of evidence} \\
\midrule

Patient outcomes (e.g. diagnostic accuracy, treatment adherence) & 
X (RCTs, cohort studies) & X & X & X & X & X & X & X \\

\midrule
Healthcare workflow efficiency & 
X (RCTs, cohort studies) & X & X & X & X & X & X & X \\

\midrule
Mortality reduction & 
X (RCTs, cohort studies) & X & X & X & X & X & X & X \\

\midrule
Quality of care improvement & 
X (RCTs, cohort studies) & X & X & X & X & X & X & X \\

\midrule
Adverse events (e.g. decision-making errors) & 
X (RCTs, cohort studies) & X & X & X & X & X & X & X \\
\bottomrule
\end{tabular}

\vspace{0.8em}
\footnotesize
\textbf{Abbreviations:} CI: Confidence Interval; GRADE: Grading of Recommendations, Assessment, Development, and Evaluation; RCT: Randomized Controlled Trial. \\[0.5em]
\textbf{Notes:}
\begin{itemize}
    \item Risk of bias: Assessed using appropriate tools for RCTs and observational studies (e.g., Cochrane Collaboration tool).
    \item Inconsistency: Refers to unexplained heterogeneity between studies, assessed via I\textsuperscript{2} and visual forest plot inspection.
    \item Indirectness: Accounts for differences in patient population, intervention, or outcomes compared to the study question.
\end{itemize}
\end{threeparttable}
\end{table}
\end{landscape}

\subsubsection*{Funding}
This research received no specific grant from any funding agency in the public, commercial or not-for-profit sectors.

\subsubsection*{Competing interests}
The authors declare no competing interests.

\subsubsection*{Authors’ contributions}
GJ: concept refinement, protocol design, PROSPERO registration, drafting. 

AB: methodological review and revisions. SP: conceptualisation, supervision, and final approval.

Guarantor: GJ.

\textbf{Ethics approval and consent to participate :} Not applicable

\textbf{Consent for publication:} All authors consent for the publication

\textbf{Availability of data and material:} Not applicable

\textbf{Acknowledgements:} None

\textbf{Patient and public involvement:} Patients or the public were not involved in the design, conduct, reporting, or dissemination plans of this research.

\bibliographystyle{plain}
\bibliography{references}

\end{document}